\documentclass[preprint,aps,tightenlines,showpacs,nofootinbib]{revtex4}
\usepackage{latexsym}
\usepackage[dvips]{graphicx}
\newcommand{\beq}{\begin{eqnarray}}
\newcommand{\eeq}{\end{eqnarray}}
\newcommand{\eq}{eqnarray}

\newcommand{\al}{{\alpha}}

\newcommand{\ci}{\cite}

\newcommand{\Ga}{{\Gamma}}
\newcommand{\ep}{{\epsilon}}

\newcommand{\de}{{\delta}}
\newcommand{\De}{\Delta}

\newcommand{\La}{{\Lambda}}
\newcommand{\m}{{\mu}}
\newcommand{\n}{{\nu}}

\newcommand{\pa}{{\partial}}
\newcommand{\no}{{\nonumber}}
\newcommand{\f}{\frac}
\newcommand{\ra}{\rightarrow}

\newcommand{\temp}{temperature }

\newcommand{\fn}{\footnote}
\newcommand{\appr}{approximation}

\begin{document}

\preprint{arXiv:0811.2685v4 [hep-th]}

\title{Smeared Hairs and Black Holes in Three-Dimensional de Sitter
Spacetime }

\author{Mu-In Park\footnote{E-mail address: muinpark@yahoo.com}}

\affiliation{ Research Institute of Physics and Chemistry, Chonbuk
National University, Chonju 561-756, Korea }

\begin{abstract}
It is known that there is no three-dimensional analog of de Sitter
black holes.
I show that the analog does exist when {\it non}-Gaussian (i.e.,
ring-type) smearings of point matter {\it hairs} are considered.
This provides a new way of constructing black hole solutions from
hairs. I find that the obtained black hole solutions are quite
different from the usual {\it large} black holes in that there are
i) large to small black hole transitions which may be considered as
{\it inverse} Hawking-Page transitions and ii) soliton-like (i.e.,
non-perturbative) behaviors. For Gaussian smearing, there is no
black hole but a {\it gravastar} solution exists.

\end{abstract}

\pacs{04.60.Kz, 04.60.-m, 04.70.Dy }

\maketitle

\newpage

\section{Introduction}

Over the years black hole solutions have played a key role in recent
developments in theoretical physics due to its unique window into
quantum gravity. In particular, the lower-dimensional black holes,
like BTZ black hole \ci{Bana:92}, has been crucial in understanding
the holographic description of asymptotically anti-de Sitter (AdS)
spacetimes\ci{Ahar:00}, other than as just a simpler analog of its
higher dimensional counterpart.

In asymptotically de Sitter (dS) spacetimes, however, the
lower-dimensional analog of black holes is not known, even though
there are solutions with the cosmological horizon
\ci{Dese:84,Park:98}. It would be desirable to have available a
lower-dimensional analog which could exhibit the key features of its
higher-dimensional counterparts. This would be important in the
holographic description of dS spacetimes also
\ci{Park:98,Stro:01,Bala:02}, as BTZ black hole did in AdS. On the
other hand, the conventional black holes are perturbative solution
in that their horizon sizes scale  with (some positive powers of)
the Newton's constant $G$.

There have been numerous works on the emergence of a new (inner)
horizon in the black hole spacetimes when the Gaussian but {\it
an}isotropic, i.e., $p_r \neq p_{\phi}$, smearing of point-matter
distributions with energy-momentum ${T^\m}_\n={\rm diag} (- \rho,
p_r, p_{\phi},...)$, in a ``self-consistent'' way \ci{Anso:08}.
However, the Gaussianity needs not be required always, though beyond
the Gaussianity has not been studied much so far: As a deformation
of $\delta$-function source, it is enough that the distribution has
a sharp peak at the origin as the characteristic size shrinks to
zero, maintaining the integration of the distribution function to be
finite so that it can be normalized to unity always. Actually, the
necessity of this kind of non-Gaussian {\it regularization} of the
$\delta$-function has been noted earlier, in a study of quantum
gravity by DeWitt \ci{Dewi:67}, though its physical origin was not
understood. More recently, Myung and Yoon have constructed a
deformed $AdS_3$ black hole, based on Rayleigh distribution which is
one of the non-Gaussian distributions \ci{Myun:08}. It is the
purpose of this paper to report that the three-dimensional analog of
dS black holes does exist when {\it non}-Gaussian smearings of point
matter {\it hairs} are considered, generally\footnote{For an early
treatment on the $dS$ black holes in higher dimensions, see
\ci{Dymn:98}.}.

 In this paper, I consider the general form of matter
distributions, which include the Gaussian, Rayleigh, and
Maxwell-Boltzmann distributions with moments $n=0, 1, 2$,
respectively. It is shown that there are black hole solutions for
all the higher-moment matter distributions, except the conventional
Gaussian one. This provides a new way of constructing black hole
solutions {\it from} hairs. It is found that the strong energy
condition is satisfied near the black holes so that the usual area
(increasing) theorem is guaranteed, like the conventional black
holes in {\it vacuum}. By demanding the area law, following the
Bekenstein's argument, and the first law of thermodynamics, the
black hole mass is identified. However, I find that the obtained
black hole solutions are quite different from those of the
conventional (large) black holes in thermodynamical properties in
that there are large to small black hole transitions which may be
considered as inverse Hawking-Page transitions. The black holes show
also soliton-like (i.e., non-perturbative) behaviors. For the
Gaussian distribution, there is no black hole but a gravastar
solution exists, instead.



\section{De Sitter black holes and
gravastars from smeared matters}

%
%
%
%
%
%

The three-dimensional Einstein gravity with a positive cosmological
constant $\La=+1/{l}^2$ is described by the action on a manifold
${\cal M}$ ( omitting some boundary terms )
\begin{\eq}
\label{EH} I_{g}=\frac{1}{16 \pi G} \int_{\cal M} d^3 x \sqrt{-g}
\left( R -2 \La~\right)+I_{\rm matter},
\end{\eq}
where $I_{\rm matter}$ is a matter action whose (microscopic)
details are not important in this paper. The equations of motion for
the metric are given by
\begin{\eq}
\label{eom} R^{\mu \nu}-\f{1}{2} g^{\m \n} R +\f{1}{{l}^2} g^{\m \n}
=8 \pi G ~T^{\m \n}_{\rm matter}
\end{\eq}
with the matter's energy-momentum tensor
\begin{\eq}
\label{T:matter}
 T^{\m \n}_{\rm matter}
 =-\f{2}{\sqrt{-g}} \f{\de I_{\rm matter}}{\de g_{\m \n}}.
\end{\eq}

In order to solve the equation (\ref{eom}), I take the static metric
ansatz
\begin{eqnarray}
\label{BTZ}
 ds^2=-N^2 (r) dt^2 +N^{-2} (r) dr^2 +r^2 d \phi ^2.
\end{eqnarray}
Here note that $g_{tt}=-g_{rr}^{-1}$ like the usual vacuum solution
\ci{Dese:84,Park:98} but this may not be true for arbitrary $T_{\m
\n}$. So, I am considering the specific matter configurations which
do not deform (\ref{BTZ}). By considering matter's energy-momentum
\begin{\eq}
\label{T:an-iso}
 {T^\m}_\n={\rm diag} (- \rho, p_r, p_{\phi}),
\end{\eq}
the Einstein equation (\ref{eom}) reads
\begin{\eq}
\f{(N^2)'}{2r}&=&-\La-8 \pi G \rho, \\
\f{(N^2)'}{2r}&=&-\La+8 \pi G p_r, \\
\f{(N^2)''}{2r}&=&-\La+8 \pi G p_{\phi},
\end{\eq}
where prime $(')$ denotes the derivative with respect to the radial
coordinate $r$. The solutions of $N^2, \rho, p_r, p_{\phi}$ are
given by
\begin{\eq}
N^2&=&-\La r^2 -16 \pi G \int^r_0 \rho r dr, \label{N^2} \\
p_r&=&-\rho, \label{p:r}\\
p_{\phi}&=&-(r \rho)',  \label{p:phi}
\end{\eq}
where I have set $N(0)=0$ in order to be agreed with the vacuum de
Sitter solution for $\rho=0$. Eqs. (\ref{p:r}) and (\ref{p:phi})
show that a non-vanishing
radial pressure $p_r$ and an-isotropic tangential pressure
$p_{\phi}=-\rho-r \rho '$ for an arbitrary matter distribution with
$\rho ' \neq 0$ are needed, respectively. Hence, the ansatz
(\ref{BTZ}), together with (\ref{T:an-iso}), determines the metric
and matter's pressures completely, in terms of the matter density
$\rho$.

Now, let me introduce the matter density as
\begin{\eq}
\label{density} \rho=A \f{r^n}{L^{n+2}} e^{-r^2/L^2},
\end{\eq}
where $L$ is a characteristic length scale of the matter
distribution and $A$ is a normalization constant. One can obtain the
Gaussian distribution for $n=0$, and non-Gaussian (i.e., ring-type)
distributions for higher moments, i.e., Rayleigh for $n=1$,
Maxwell-Boltzman for $n=2$, etc. Then, by plugging (\ref{density})
into (\ref{N^2}), one can easily find
\begin{\eq}
 N^2&=&-\La r^2 -8 \pi G A ~\gamma \left(\f{n}{2}+1,
\f{r^2}{L^2}\right) \no \\  &=&-\La r^2 -8 \pi G A ~\left[\Gamma
\left(\f{n}{2}+1 \right)-\Gamma \left(\f{n}{2}+1,
\f{r^2}{L^2}\right)\right], \label{dSBH}
\end{\eq}
where $\gamma (\f{n}{2}+1, x^2)=\int^{x^2}_0 t^{n/2} e^{-t} dt$,
$\Gamma (\f{n}{2}+1, x^2)=\int_{x^2}^{\infty} t^{n/2} e^{-t}
dt=\Gamma (\f{n}{2}+1)-\gamma (\f{n}{2}+1, x^2)$ are the incomplete
lower and upper Gamma functions, respectively.

What we know is that the Einstein equation is {\it uniquely} solved
by the $dS_3$ solution \ci{Dese:84,Park:98}
\begin{\eq}
\label{dS_3}
 N^2_{dS_3}=-\f{r^2}{l^2} +8 G m
\end{\eq}
in the vacuum, i.e., $L \ra 0$, limit and there is a cosmological
horizon at $r_+=l \sqrt{8Gm}$ with the de Sitter mass $m$.
From this boundary condition, which distinguishes the matter
distribution for the smearing of point sources with the {\it
classical} matter distributions, one obtains
\begin{\eq}
\label{A}
 A=-\f{m}{ \pi \Gamma \left(\f{n}{2}+1\right)}
\end{\eq}
and then (\ref{dSBH})  reads
\begin{\eq}
\label{N^2:final}
 N^2=-\f{r^2}{l^2} +8 G m \left[ 1- \f{\Gamma
\left(\f{n}{2}+1, \f{r^2}{L^2}\right)}{\Gamma
\left(\f{n}{2}+1\right)}\right]
\end{\eq}
with $\La=1/l^2$. It is easy to see that this reduces to the $dS_3$
metric with (\ref{dS_3}) from the fact that the last term in $[~ ]$
vanishes in the $L \ra 0$ limit. In the presence of the smeared
matters the cosmological horizon of the (vacuum) $dS_3$ solution
(\ref{dS_3}) would be shifted also. Now, in order to see whether
there exists a black hole horizon, I need to know whether $N^2=0$
has interior roots, other than the cosmological horizons. The (black
hole and cosmological) horizons, if exist, satisfy\footnote{In the
{\it flat} space with a vanishing cosmological constant, i.e., $l
\ra \infty$, one rather solves $N^2=(1-4G \bar{m})^2 \left[ 1-
{\Gamma \left(\f{n}{2}+1, \f{r^2}{L^2}\right)}/{\Gamma
\left(\f{n}{2}+1\right)}\right]=0$ with a particle's mass $\bar{m}$.
But it is easy to see that the horizon does not occur since ${\Gamma
\left(\f{n}{2}+1, \f{r^2}{L^2}\right)} <{\Gamma
\left(\f{n}{2}+1\right)}$ always for $r>0$. }
\begin{\eq}
\label{horizon}
 \hat{r}_{\pm}^2 = 8G m l^2 \left[ 1- \f{\Gamma
\left(\f{n}{2}+1, \f{\hat{r}_{\pm}^2}{L^2}\right)}{\Gamma
\left(\f{n}{2}+1\right)}\right]
\end{\eq}
but it is hard to solve this analytically.

An easier way to find the existence of the (interior) black hole
horizons is to consider the, so-called, {\it Nariai} limit, where
the black hole horizon $\hat{r}_-$ and the cosmological horizon
$\hat{r}_+$ meet, i.e., $\hat{r}_{-}=\hat{r}_{+}\equiv {\hat r}_{\rm
Nar}$, at which $(N^2)'=0$ as well as $N^2=0$. If there exists a
{\it positive} solution for the {\it Nariai radius} $\hat{r}_{\rm
Nar}$, the black hole horizon $\hat{r}_-$ also exist, as well as the
cosmological horizon $\hat{r}_+$, since $\hat{r}_{-} \leq
\hat{r}_{\rm Nar} \leq \hat{r}_{+}$. From $(N^2)'=0$, one has
\begin{\eq}
r=\f{8 Gm l^2 \left( \f{r}{L} \right)^{n+1}}{ \Gamma
\left(\f{n}{2}+1\right) L} e^{-\f{r^2}{L^2}}.
\end{\eq}
By plugging this into the left hand side of (\ref{horizon}), with an
identification of $r=\hat{r}_{\pm}=\hat{r}_{\rm Nar}$, one has an
algebraic equation
\begin{\eq}
f(\hat{r}_{\rm Nar};L) \equiv \left( \f{\hat{r}_{\rm Nar}}{L}
\right)^{n+2} e^{-\f{\hat{r}_{\rm Nar}^2}{L^2}}-\gamma
\left(\f{n}{2}+1, \f{\hat{r}_{\rm Nar}^2}{L^2} \right)=0,
\end{\eq}
which does not depend on the dimensionful parameters of $G, ~m$, and
$l^2$, but depends only on the dimensionless parameter $x_{\rm Nar}
\equiv \hat{r}_{\rm Nar }/L$, for each $n$. So, the existence
problem of the $dS_3$ black holes reduces to a purely mathematical
problem of finding the number $x_{\rm Nar (n)}$ and let me call this
{\it Nariai number}, for convenience. The explicit formula for the
Nariai numbers is not known but one can find them numerically by
plotting $f(\hat{r}_{\rm Nar};L)$ as in Fig.1, which reads $x_{\rm
Nar (0)}=0,~ x_{\rm Nar (1)} \approx 0.96786, ~ x_{\rm Nar (2)}
\approx 1.33914, ~ x_{\rm Nar (3)} \approx 1.61363, ~ x_{\rm Nar
(4)} \approx 1.83947$, etc.

\begin{figure}
\includegraphics[width=12cm,keepaspectratio]{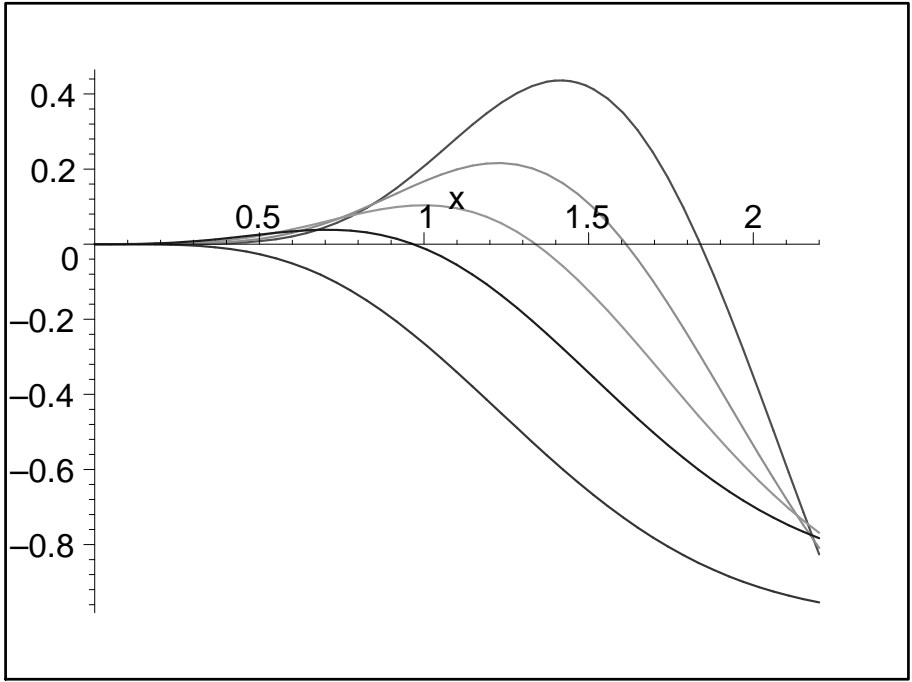}
\caption{Plots of $f(\hat{r}_{\rm Nar};L)$ vs. $x_{\rm
Nar}=\hat{r}_{\rm Nar}/L$ for various $n$. The Nariai numbers
$x_{\rm Nar}$ are the intersection points of $f(\hat{r}_{\rm
Nar};L)=0$, other than the trivial one at $x_{\rm Nar}=0$. Nariai
number increases indefinitely as $n$ is increased. Numerically, it
reads $x_{\rm Nar (0)}=0,~ x_{\rm Nar (1)} \approx 0.96786, ~ x_{\rm
Nar (2)} \approx 1.33914, ~ x_{\rm Nar (3)} \approx 1.61363, ~
x_{\rm Nar (4)} \approx 1.83947$, etc. (left to right).}
\label{fig:Nariai}
\end{figure}

This result shows that there exist (three-dimensional) de Sitter
black holes only for the non-Gaussian distributions, i.e., $n \geq
1$. On the other hand, for the Gaussian distributions the solution
correspond to a ``gravastar'', a compact self-gravitating object
without horizons \ci{Mazu:01}.

\section{Properties of de Sitter black holes}

In order to study the physical properties of de Sitter black holes,
I need to know the detailed form of the horizon $\hat{r}_-$.
However, since (\ref{horizon}) can not be solved analytically, I
consider the {\it perturbative} solution near the origin, by
demanding $\hat{r}_-$ is very small; this is a reasonable assumption
for small $L$ since $\hat{r}_-$ should be disappeared as $L$ goes to
zero.

By expanding (\ref{N^2:final}) near $r \simeq 0$, neglecting higher-
order terms 
\begin{\eq}
N^2 = -\f{r^2}{l^2} \left[ 1-\f{16 G m l^2}{(n+2) \Gamma
\left(\f{n}{2}+1\right)}\f{r^n}{L^{n+2}} + {\cal
O}\left(\f{r^{n+2}}{L^{n+2}}\right) \right],
\end{\eq}
one finds the black hole horizon as
\begin{\eq}
\hat{r}_{-} \simeq \left(\f{(n+2) \Gamma \left(\f{n}{2}+1\right)}{16
G m l^2} \right)^{\f{1}{n}} L^{1+\f{2}{n}}.
\end{\eq}
The black hole horizon is proportional to (positive powers of) $L$
and so becomes very small as $L \ra 0$, which provides a validity
criteria of the approximation. In contrast to the (outer)
cosmological horizon, the size of black hole is ``inversely''
proportional to the {\it initial} $dS_3$ mass $m$ and $l$.
Furthermore, the black holes are {\it soliton-like} (i.e.,
non-perturbative) objects since their (horizon) sizes dominate in
weaker couplings of the Newton's constant $G$.

The Hawking temperature is obtained as
\begin{\eq}
\label{T:H}
 T_{H}= \left.\f{ \hbar \kappa}{2 \pi}\right|_{\hat{r}_{\pm}} =\f{
\hbar \hat{r}_{\pm} }{2 \pi l^2} \left| 1- x_{\pm}^{n+2} \f{e^{-
x^2_{\pm}}}{\gamma \left(\f{n}{2}+1, x^2_{\pm} \right)}\right|
\end{\eq}
with the (positive)\footnote{In dS space, there is a subtlety in
defining the temperature, associated with the definition of the
mass. But here I take the usual convention with the positive surface
gravity and temperature
 \ci{Gibb:77,Park:08}.} surface gravity function $\kappa=|\pa N^2 /2 \pa
r|$ and $x_{\pm}=\hat{r}_{\pm}/L$. In deriving (\ref{T:H}), I have
expressed the initial $dS_3$ mass $m$ in terms of $\hat{r}_{\pm}$,
from (\ref{horizon}). Note that the temperature (\ref{T:H}) is an
exact form in $\hat{r}_{\pm}$, though we do not know the exact
algebraic form of $\hat{r}_{\pm}$ in terms of $m$ and $n$. Some of
the plots for $n=0 \sim 4$ are shown in Fig.2 (solid lines). For the
Gaussian case ($n=0$), there is {\it no} zero-temperature cusp: This
is due to the absence of a black hole horizon $\hat{r}_-$ and
(\ref{T:H}) is the Hawking temperature for the cosmological horizon
$\hat{r}_+$ only. For the non-Gaussian cases ($n \geq 1$), there are
two branches of temperature curves with a zero-temperature cusp at
the intersections: The left-hand side curves represent the black
hole temperatures for the horizon $\hat{r}_-$ and the right-hand
side curves represent those of cosmological horizons $\hat{r}_+$.
The systems are in thermal equilibrium with zero temperature at the
Nariai limit, like the usual higher-dimensional {\it dS} black hole
systems \ci{Gibb:77}.\fn{In AdS or flat cases, this corresponds to
the extremal black holes \ci{Myun:08,Myun:07}.} But it is remarkable
that there exists an upper bound of the black hole temperature,
which will be discussed further below. On the other hand, the
temperature of the cosmological horizon becomes lower, due to the
existence of zero temperature at the Nariai limit\footnote{This is
in sharply contrast to the {\it G(E)UP} modifications though the
existence of the minimal sizes are the same \ci{Park:08b}.}.

\begin{figure}
\includegraphics[width=12cm,keepaspectratio]{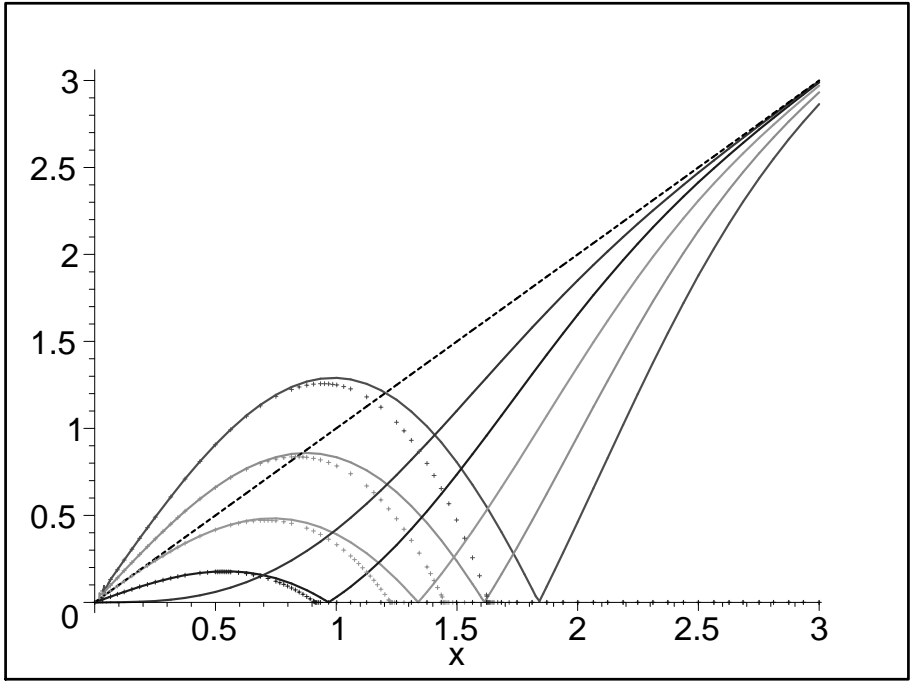}
\caption{Plots of Hawking temperature vs. $x= \hat{r}_{\pm}/L$ for
$n=0 \sim 4$ (bottom to top in the small (left) branches and top to
bottom in the large (right) branches).  The solid lines denote the
exact formula (\ref{T:H}) and the cross lines denote the small black
hole approximations in (\ref{T:approx}). The dotted line denotes the
vacuum $dS_3$ (i.e., $\rho =p=0$). I have chosen $ \hbar/2 \pi l^2
=L \equiv 1$.} \label{fig:Temp}
\end{figure}

In order to study the black hole thermodynamics, let me consider the
{\it small} black hole case, i.e., small $\hat{r}_-$, again which
reduces the Hawking temperature as
\begin{\eq}
\label{T:approx}
 T_{H}&=& \f{ n \hbar }{4  \pi l^2} \hat{r}_{-}
\left[ 1- \f{2(n+2)}{n (n+4)} \left(\f{\hat{r}_{-}}{L} \right)^2 +
{\cal O} \left(\left(\f{\hat{r}_{-}}{L} \right)^4\right)\right].
\end{\eq}
Note that the \temp vanishes for the Gaussian ($n=0$) case, as
expected, and those for non-Gaussian ($n \geq 1$) cases are plotted
in Fig.2 (cross lines), in comparison with the exact curves (solid
lines). This shows that there are good agreements for the first
several non-Gaussianities. From this approximate \temp formula
(\ref{T:approx}), one can find the approximate formula for ``small''
Nariai numbers $x_{\rm Nar }=\hat{r}_-/L$ where $T_H$ vanishes:
\begin{\eq}
\label{Nariai:approx}
 x_{\rm Nar (n)} \simeq \sqrt{\f{n(n+4)}{2 (n+2)}}.
\end{\eq}
The comparisons with the numerical results are shown in Table 1.
Note that the discrepancy increase as $x_{\rm Nar (n)}$ increases as
it would be.

\begin{table}
\begin{center}
\begin{tabular}{|c|c|c|} \hline
n& $x_{\rm Nar (n)}$, Numerical  & Approximation \\ \hline \hline 0 & 0 & 0 \\
\hline 1 & 0.96786 & 0.91287 (-5.7 \%)
\\ \hline 2 & 1.33914 & 1.22475 (-8.5 \%) \\ \hline 3 & 1.61363 & 1.44914 (-10.2 \%)\\
\hline 4 & 1.83947 & 1.63299 (-11.2 \%)\\ \hline
\end{tabular}
\end{center}
\caption{Comparisons of numerical values of the Nariai numbers
$x_{\rm Nar (n)}$ and their first approximations of
(\ref{Nariai:approx}). The values in the brackets denote the
discrepancies with the numerical values.} \label{Table:Nariai}
\end{table}

For the black hole entropy, there is no canonical derivation, like
the Euclidean action approach. But it does not seem that these black
holes violate the Bekenstein's area {\it law}, i.e., entropy being
``linearly'' proportional to the horizon area \ci{Beke:73}, since
the Hawking's area (increasing) {\it theorem} is guaranteed due to
the strong energy conditions, as will be discussed in detail below.
In other words, I {\it demand} that the black hole entropy as
\begin{\eq}
S 
\equiv \al \f{2 \pi \hat{r}_{-}}{4 \hbar G}.
\end{\eq}
But here, the (positive) coefficient $\al$ is not fixed to $1$ as in
the Bekenstein-Hawking entropy for the conventional {\it large}
black holes \ci{Hawk:71}, similarly to the so-called ``small black
holes'' in higher curvature supergravities \ci{Cai:08}; this would
be determined from other independent analysis.

Demanding the first law of thermodynamics $dM =T_H dS 
=\f{2 \pi \al } {4 \hbar G} d \hat{r}_{-}T_H$ yields the black hole
mass, if monotonically increasing in $\hat{r}_-$, as
\begin{\eq}
\label{M}
 M(\hat{r}_{-})&=& \f{2 \pi \al}{4G \hbar}
\int^{\hat{r}_{-}}_{0} d
\hat{r}_{-} T_{H} (\hat{r}_{-}) \no \\
&=&\f{ \al L^2}{4G l^2} \int^{\hat{r}_{-}/L}_{0} d x ~ x \left| 1-
x^{n+2} \f{e^{- x^2}}{\gamma \left(\f{n}{2}+1, x^2 \right)}\right|,
\end{\eq}
where I have set $M(0)=0$ in order to be agreed with the
conventional vacuum $dS_3$ solution without black holes, i.e.,
$\hat{r}_{-}=0$. The analytic integration of (\ref{M}) is not
available. But for the small black hole \appr,  (\ref{M}) reads
\begin{\eq}
M(\hat{r}_{-}) =\f{ \al}{16 G l^2} \hat{r}_{-}^2 \left[
1-\f{n+2}{n(n+4)} \left( \f{\hat{r}_{-}}{L}\right)^2  +{\cal O}
\left( \left( \f{\hat{r}_{-}}{L}\right)^4 \right) \right].
\end{\eq}
In this approximate formula, the mass has maxima at the Nariai
radius where the \temp vanishes by definition, from $dM/d \hat{r}_-=
(2 \pi \al/G \hbar) T_{H}=0$. Since $\hat{r}_-$ is less than or
equal to the Nariai radius, the black hole mass is always
monotonically increasing in the allowed regions, consistently with
the assumption for (\ref{M}).

The heat capacity
$
C={d M}/{d T_{H}}
$
is given by
\begin{\eq}
C&=&\f{ \al \pi L}{2 G} T_{H} \left( \f{d T_H}{d x} \right)^{-1}, \\
\f{d T_H}{d x}&=&\f{ \hbar L}{2 \pi l^2} \f{1}{{\gamma
\left(\f{n}{2}+1, x^2 \right)}^2} \left[ {\gamma \left(\f{n}{2}+1,
x^2 \right)}^2 -(n+3-2 x^2) x^{n+2} e^{-x^2} \gamma
\left(\f{n}{2}+1, x^2 \right) + x^{2n +3} e^{-2x^2} \right] \no
\end{\eq}
with $x \equiv \hat{r}_-/L$. There is an infinite
discontinuity\footnote{The rotating black {\it M2}-branes have the
same critical exponent  but its heat capacity is always positive
\ci{Cai:99}.} in the heat capacity $C \sim \ep |T-T_c|^{-1/2}$ [
$\ep \equiv {\rm sign}(\hat{r}_c-\hat{r})$ ]  at the location of the
maximum temperature $T_c$ and the critical horizon radius
$\hat{r}_c$ since $T-T_c \sim (\hat{r}-\hat{r}_c)^2$.
The critical location is obtained approximately as
\begin{\eq}
x_c \equiv \hat{r}_c/L=\sqrt{\f{n(n+4)}{6 (n+2)}}
\end{\eq}
from
\begin{\eq}
\f{d T_H}{d x}=\f{n  \hbar L}{4 \pi l^2} \left[ 1- \f{6(n+2)}{n
(n+4)} x^2+ {\cal O}( x^4 ) \right]
\end{\eq}
for the small black holes. As for the black holes radiate, $T_H$
decreases $(C>0)$ for the smaller black holes with
$\hat{r}_-<\hat{r}_c$ (i.e., $x<x_c$) but $T_H$ increases ($C<0$)
for the larger black holes with $\hat{r}_->\hat{r}_c$ (i.e.,
$x>x_c$). So, there are transitions between the {\it locally}
thermodynamically stable ($C>0$) and unstable ($C<0$) phases but it
is peculiar that the smaller ones are more stable, in contrast to
the three-dimensional {\it Kerr-de Sitter} \ci{Park:98} or
higher-dimensional $AdS$ black holes in {\it vacuum} \ci{Hawk:83}.

Generally, the ``local'' thermodynamic instability does not
necessarily imply the ``global'' instability, i.e., unstable to
decay into {\it globally} favored states via quantum tunneling
\ci{Pres:00}. In the canonical ensemble with a fixed temperature,
the global (in)stability is governed by the (Helmholtz) free energy
$F=M-T_H S$ in such a way that the free energy is minimized,
globally. The numerical plots of the free energy vs. Hawking
temperature are shown in Fig. 3. It is important to note that there
are two, upper and lower, branches for the large and small black
holes, respectively, with $\De F \equiv F_{\rm small}-F_{\rm large}
<0$ at the {\it same} temperature and they meet exactly at the point
where $T_H$ is maximum\footnote{This shows a second-order phase
transition since $F$ and $dF/dT=-S$ are continuous but only $d^2
F/dT^2=-C/T$ is discontinuous at the critical point.}. This means
that the region of local (in)stability coincides ``exactly'' with
that of global (in)stability and the global transitions via
tunnelings from large to small black holes corresponds to
``inverse'' Hawking-Page transitions \ci{Hawk:83}\footnote{In
charged AdS black holes, similar phenomena can happen, for a range
of temperature, in the canonical ensemble with a (fixed) charge $q <
q_{\rm crit}$ \ci{Cham:99}. (See the branches 1 and 2 in Fig. 5.)
Recently, this has been argued in the higher-dimensional regular
black holes \ci{Myun:07,Myun:07b}, where the lower bounds of the
horizons exist at the extremal black holes, but the status of the
ADM mass or the first law of thermodynamics for the regular black
holes remains unclear. }, with tunneling amplitudes $\Ga \sim A
e^{-\De I_E}$ and $\De I_E \approx -\De F/T_H
>0$ ($A$ is some determinant and $I_E$ is the {\it on-shell} Euclidean action).
There is no lower bound in the size of small black holes but rather
an upper bound at the Nariai limit. As one increases the size of the
small black hole, i.e., increasing the characteristic scale $L$,
there are more smearings of the matter hairs {\it outside} the
(inner, black-hole) horizon  until the critical size $\hat{r}_c$
when the maximum temperature $T_c$ is reached. Further increase of
the size beyond the critical size produces tunneling decays into the
smaller ones with the same temperature, by {\it re}-absorbing the
excess matter hairs on the one hand and more rapid Hawking
radiations due to $C<0$ on the other hand. I note also that the
transitions from small black holes with higher moments ($n'$) into
large or other small black holes with lower moments ($n$) are also
possible, i.e., $ F_{\rm large,~small (n)}-F_{\rm small(n')} <0
~(n<n')$, when the temperatures have overlapping regions.

\begin{figure}
\includegraphics[width=12cm,keepaspectratio]{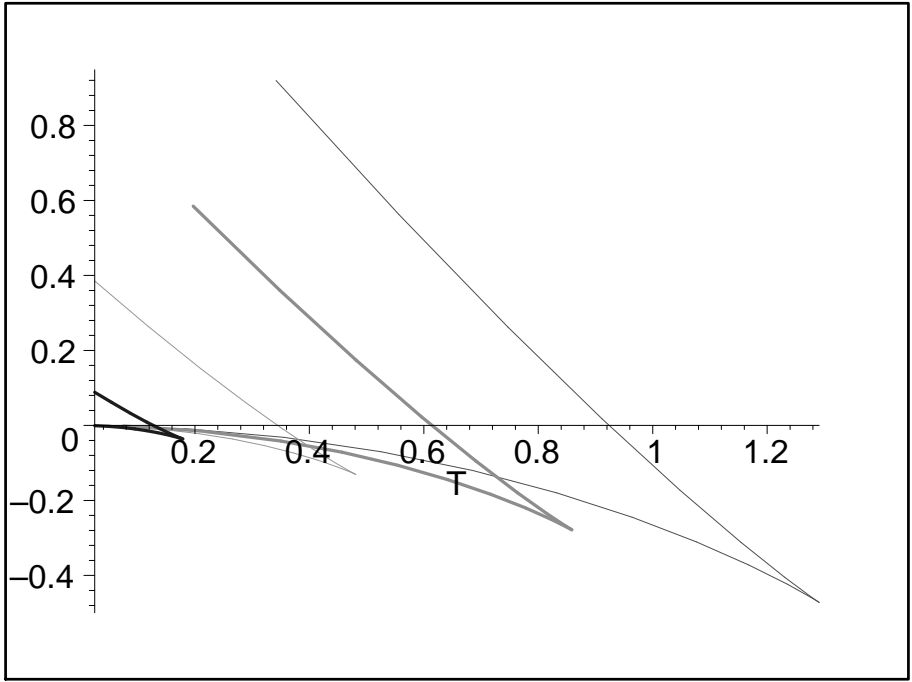}
\caption{Plots of free energy $F$ vs. Hawking temperature for $n=1
\sim 4$ (left to right) ($ \hbar/2 \pi l^2 =L \equiv 1$). For each
curve with a given $n$, there are two (upper and lower) branches
which meet at the maximum temperature $T_c$. The upper (lower)
branch represents the large (small) black hole, giving $ F_{\rm
small}-F_{\rm large} <0$. } \label{fig:Temp}
\end{figure}

\section{Energy conditions}

From the normalization of (\ref{A}), one finds the energy density of
smeared (point) matters as\footnote{This can be also written as
$\rho= -\f{m}{2 \pi L} \f{\Gamma \left(\f{n+1}{2}\right)}{\Gamma
\left(\f{n}{2}+1\right)} \hat{\de}_{n}(r)$ with the non-Gaussian
smearing of $\de$-function, $\hat{\de}_{n}(r)=\f{2r^n}{\Gamma
\left(\f{n+1}{2}\right) L^{n+1}} e^{-r^2/L^2}$, which satisfies
$\lim_{L \ra 0} \hat{\de}_{n}(r)=\de (r)$. Sometimes, it has been
said that the black holes ``degenerate'' to a conical singularity at
the origin, by Bousso, et. al. \ci{Bous:02}. Their idea is
``materialized'' in my construction. }
\begin{\eq}
\label{density:normalized} \rho= -\f{m}{ \pi \Gamma
\left(\f{n}{2}+1\right)} \f{r^n}{L^{n+2}} e^{-r^2/L^2}.
\end{\eq}
For the {\it initially} $dS_3$ space, one has a positive mass $m>0$,
the matter's energy density $\rho$ is negative, and so the weak and
dominant energy conditions are violated in the {\it whole} space.
However, less restrictive but more important conditions in the black
hole dynamics, like the strong or null energy condition can be
satisfied.

To see this, let me consider $\rho+p_i,~\rho+ \sum_i p_i~(i=r,
\phi)$ which are obtained as
\begin{\eq}
\label{rho:normalized}
 \rho+p_{\phi}&=&-r \rho '=\f{m}{ \pi \Gamma
\left(\f{n}{2}+1\right)
L^{n+2}}\left(n -\f{2 r^2}{L^2} \right) {r^n} e^{-r^2/L^2}, \\
\rho+ \sum_i p_i &=& p_{\phi}=\f{m}{ \pi \Gamma
\left(\f{n}{2}+1\right) L^{n+2}}\left(n+1 -\f{2 r^2}{L^2} \right)
{r^n} e^{-r^2/L^2},
\end{\eq}
where I have used $ \rho+p_{r}=0$ from (\ref{p:r}), (\ref{p:phi}),
and (\ref{density:normalized}). Then one finds that $\rho+p_{\phi}
\geq 0$ for $\f{r}{L} \leq \sqrt{\f{n}{2}}$, whereas $\rho+ \sum_i
p_i \geq 0$ for $\f{r}{L} \leq \sqrt{\f{(n+1)}{2}}$. Hence, it is
found that the strong energy condition (SEC), which includes the
null energy condition (NEC) \ci{Pois:04} is satisfied when $\f{r}{L}
\leq x_{\rm sec} \equiv \sqrt{\f{n}{2}}$. In order that the black
hole has the required behaviors like the increasing horizon area for
the accretion of the surrounding matters, I need to require
$\f{\hat{r}_{-}}{L} <x_{\rm sec}$. This yields
\begin{\eq}
\label{SEC}
 \f{8G m l^2}{L^2} > \f{ \f{n}{2} \Gamma
\left(\f{n}{2}+1\right) }{\gamma \left(\f{n}{2}+1, \f{n}{2} \right)}
\end{\eq}
from $N^2|_{x_{\rm sec}} >0$. For a good \appr, up to $n \approx
30$, the right hand side of (\ref{SEC}) converges as `$2.345 + 1.08
n$' and so (\ref{SEC}) can be written approximately as
\begin{\eq}
m \geq \f{L^2}{8 Gl^2} (2.345 + 1.08 n).
\end{\eq}
Moreover, for small black holes, it is easy to see that $x_{\rm
sec}$, which is  actually the maximum point of the density $\rho$,
i.e., $\rho ' |_{x_{\rm sec}}=0$, is smaller than the Nariai radius,
i.e., $x_{\rm sec} =\sqrt{\f{n}{2}} < x_{\rm Nar} \approx
\sqrt{\f{n}{2} (\f{n+4}{n+2} )}$, which implies that there is a
region where the SEC and NEC are violated, in between $x_{\rm sec}$
and $x_{\rm Nar}$. So, in this case, the cosmological horizon would
not satisfy the Hawking's area (increasing) theorem. However, the
black hole would satisfy the area theorem well. For the gravastar
case ($n=0$), there is no black hole horizon and the condition
(\ref{SEC}) is trivially satisfied, which means that the SEC and WEC
are trivially satisfied as far as $\f{r}{L} \leq x_{\rm sec}$.

\section{Discussion}

I have studied the $dS_3$ black holes for the non-Gaussian and
gravastars for Gaussian smearings of point-matter sources or hairs.
I have studied the particular class of metrices satisfying
$g_{tt}=-g_{rr}^{-1}$ for simplicity. It would be a challenging
problem to generalize this construction to include the ``rotating''
black holes and its associated {\it horizon smearings} which have
been perturbatively studied in non-commutative BTZ black hole
showing the interior black hole {\it inside} the inner BTZ black
hole horizon, as well \ci{Kim:08}. On the other hand, it seems that
the smeared sources can be realized in the non-commutative solitons
\ci{Gopa:00,ICho:08}\footnote{I thank H. S. Yang for pointing out
this possibility.}. It would be quite interesting to study the {\it
gravitating} non-commutative solitons and their associated black
hole spacetimes.

From the obtained $dS_3$ black holes, it is straightforward to
construct four-dimensional (A)dS black strings which share the
thermodynamical properties of $dS_3$ black holes \ci{Park:toappear}.
According to Gubser-Mitra conjecture \ci{Gubs:00}, the large black
strings with the heat capacity $C_{b.s.} \sim C <0$ would then be
unstable under gravitational perturbations. This system provides an
interesting test of the conjecture in four dimensions. Moreover,
this provides a challenging problem of the final states of black
string instability, i.e., whether it be the thinner black strings
which are favored by the lower values of free energies or the
four-dimensional Schwarzschild-de Sitter black hole, or in between
them.

I have shown that finding the Nariai radius involves an interesting,
purely mathematical, algebraic equation which is solved by the
Nariai numbers $x_{\rm Nar (n)} $. I have found an approximate
formula of $x_{\rm Nar (n)} $ for small black holes. It would be a
challenging mathematical problem to find an improved or exact
formula.

\section*{Acknowledgments}

This work was
supported by the Korea Research Foundation Grant funded by Korea
Government(MOEHRD) (KRF-2007-359-C00011).

\newcommand{\J}[4]{#1 {\bf #2} #3 (#4)}
\newcommand{\andJ}[3]{{\bf #1} (#2) #3}
\newcommand{\AP}{Ann. Phys. (N.Y.)}
\newcommand{\MPL}{Mod. Phys. Lett.}
\newcommand{\NP}{Nucl. Phys.}
\newcommand{\PL}{Phys. Lett.}
\newcommand{\PR}{Phys. Rev. D}
\newcommand{\PRL}{Phys. Rev. Lett.}
\newcommand{\PTP}{Prog. Theor. Phys.}
\newcommand{\hep}[1]{ hep-th/{#1}}
\newcommand{\hepp}[1]{ hep-ph/{#1}}
\newcommand{\hepg}[1]{ gr-qc/{#1}}
\newcommand{\bi}{ \bibitem}

\end{document}